\documentclass[runningheads]{llncs}
\usepackage[T1]{fontenc}
\usepackage{graphicx}
\usepackage{subcaption}

\begin{document}
\title{A User-Centric Analysis of Explainability\\in AI-Based Medical Image Diagnosis}
\titlerunning{User-Centric Analysis of Explainability in AI-Based Medical Image Diagnosis}

%

\author{Julia Wagner \and
Tim Schlippe\orcidID{0000-0002-9462-8610}}

\authorrunning{J. Wagner and T. Schlippe}
%
\institute{IU International University of Applied Sciences, Germany\\ 
\email{tim.schlippe@iu.org}}

\maketitle              

\vspace{-0.3cm}

\begin{abstract}
In recent years, AI systems in the medical domain have advanced significantly. However, despite outperforming humans, they are rarely used in practice since it is often not clear how they make their decisions. Optimal explanation and visualization of the decision process are often lacking. Therefore, we conducted a comparative user-centric analysis of the latest state-of-the-art textual, visual and multimodal explainable artificial intelligence (XAI) methods for medical image diagnosis. Our survey of 33~physicians showed that 88\% \textit{agree} that it is important that AI explains the diagnosis---64\% even strongly agree. A combination of \textit{bounding box} and \textit{report} is rated better than the other tested XAI methods in the evaluated aspects \textit{understandability}, \textit{completeness}, \textit{speed}, and \textit{applicability}. We even tested the potential negative impact of false AI-based medical image diagnoses and found that 50\% of the participants trusted false AI diagnoses over all tested XAI methods.

\vspace{-0.3cm}

\keywords{XAI \and visualization \and medical image analysis \and artificial intelligence \and explainable AI \and explainability \and user-centric analysis.}
\end{abstract}
\vspace{-0.7cm}

\section{Introduction}
\label{sec:intro}

\vspace{-0.1cm}

AI holds great promise for revolutionizing healthcare, addressing challenges such as physician shortages, heavy workloads, and demographic shifts. The World Health Organization reported a global deficit of 4.3 million healthcare workers in 2006, projected to reach 12.9 million by 2035~\cite{WHO2009}. 
AI’s ability to rapidly learn and scale addresses these issues, offering improvements in both quality and efficiency. 
AI applications span prevention, diagnosis, therapy, and aftercare. Medical image diagnosis is especially promising. 
Despite advancements, adoption remains limited due to the 'black box' nature of deep learning algorithms, which undermines trust~\cite{Adadi:2018}. Transparency through XAI is critical for integrating these systems into high-risk clinical settings. 

Consequently, our paper investigates textual, visual, and combined XAI methods for medical image diagnosis in a user-centric evaluation with 33 physicians. Our contributions include: (1)~evaluating state-of-the-art XAI methods; (2)~offering a detailed, example-based assessment beyond automatic evaluation methods; (3)~analyzing how incorrect AI diagnoses impact physicians’ decision-making.

\vspace{-0.3cm}

\section{Related Work}
\label{Related Work}

\vspace{-0.2cm}


\cite{VANDERVELDEN2022102470} categorizes 221~XAI papers in the medical domain into visual, textual, and example-based methods. Most papers focus on visual XAI methods, which highlight key areas of an image influencing decisions through \textit{saliency} or \textit{heatmap} visualizations~\cite{app122211750}. Textual XAI methods provide explanations in \textit{reports} and include visual question answering with \textit{chatbots} that generate responses based on visual content and dialogue history~\cite{borys2023explainable,s23020634}. 
Example-based XAI methods offer clarity by providing similar instances to the input data~\cite{s23020634}. 
Related work emphasizes the need for user-centered evaluations and academic-clinical collaboration~\cite{VANDERVELDEN2022102470}, but such analyses in medical image XAI remain limited. \cite{prentzas2023explainable} found studies averaging only 8.3~participants, with a maximum of 21. Our review identified two larger studies: \cite{make3030037} surveyed 60 participants on visual \textit{heatmap} XAI methods, not all medical professionals, while \cite{Xie:2020} involved 77~medical professionals in chest X-ray analysis. 
The only work which evaluates a multimodal combination of XAI methods is~\cite{gale2018producing}. However, in this work only five physicians evaluated one textual and one visual XAI method individually plus their multimodal combination. 
Consequently, our goal was to conduct a user study with at least 30~medical professionals to compare various visual, textual, and multimodal XAI methods. 

\vspace{-0.6cm}

\begin{figure}[h!]
\centering
\begin{subfigure}{0.35\textwidth}
    \includegraphics[width=\linewidth]{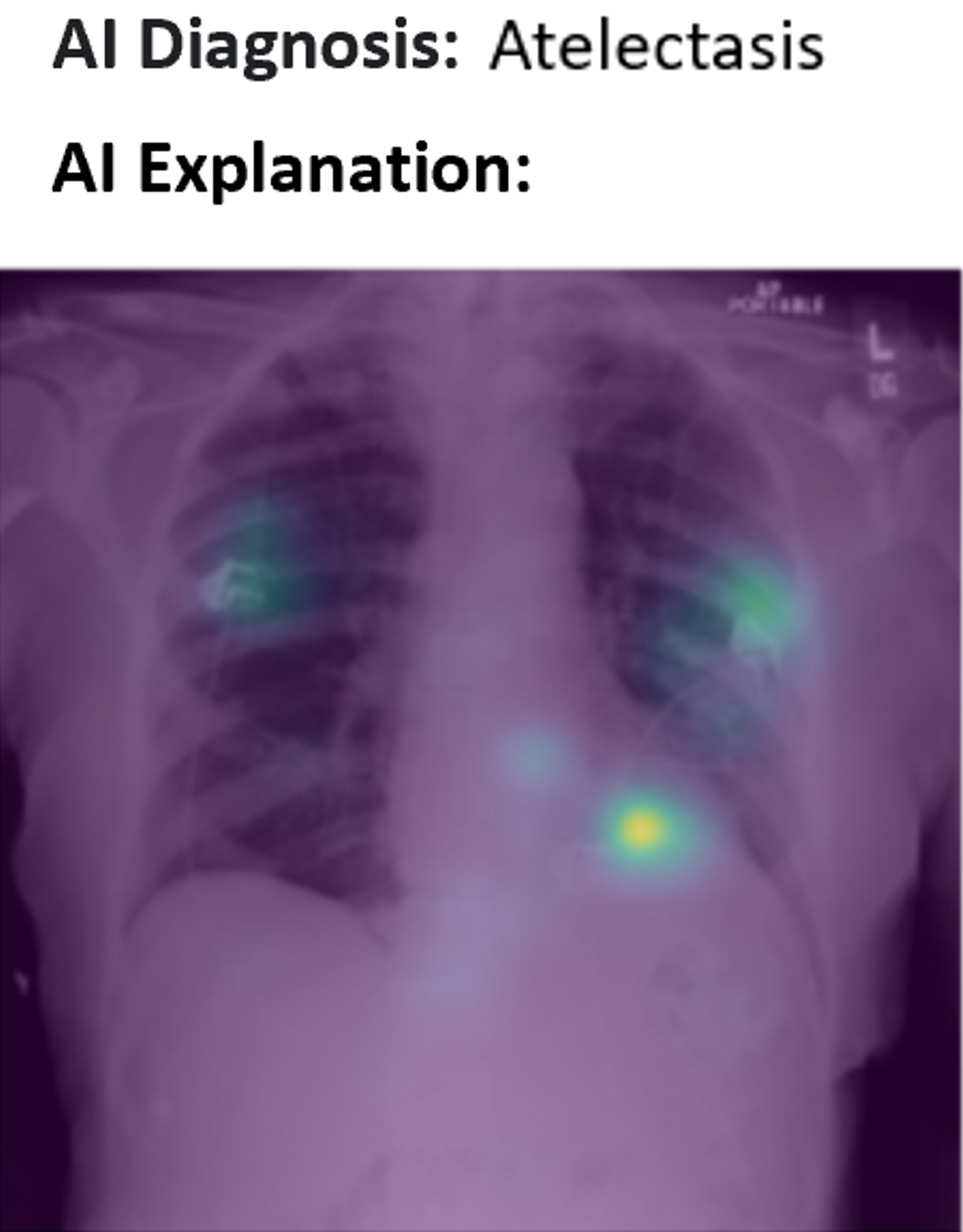}
    \vspace{-0.3cm}
    \caption{Visual XAI: \textit{Heatmap}.}
    \label{Heatmap}
\end{subfigure}\hfill
    \vspace{-0.1cm}
\begin{subfigure}{0.36\textwidth}
    \includegraphics[width=\linewidth]{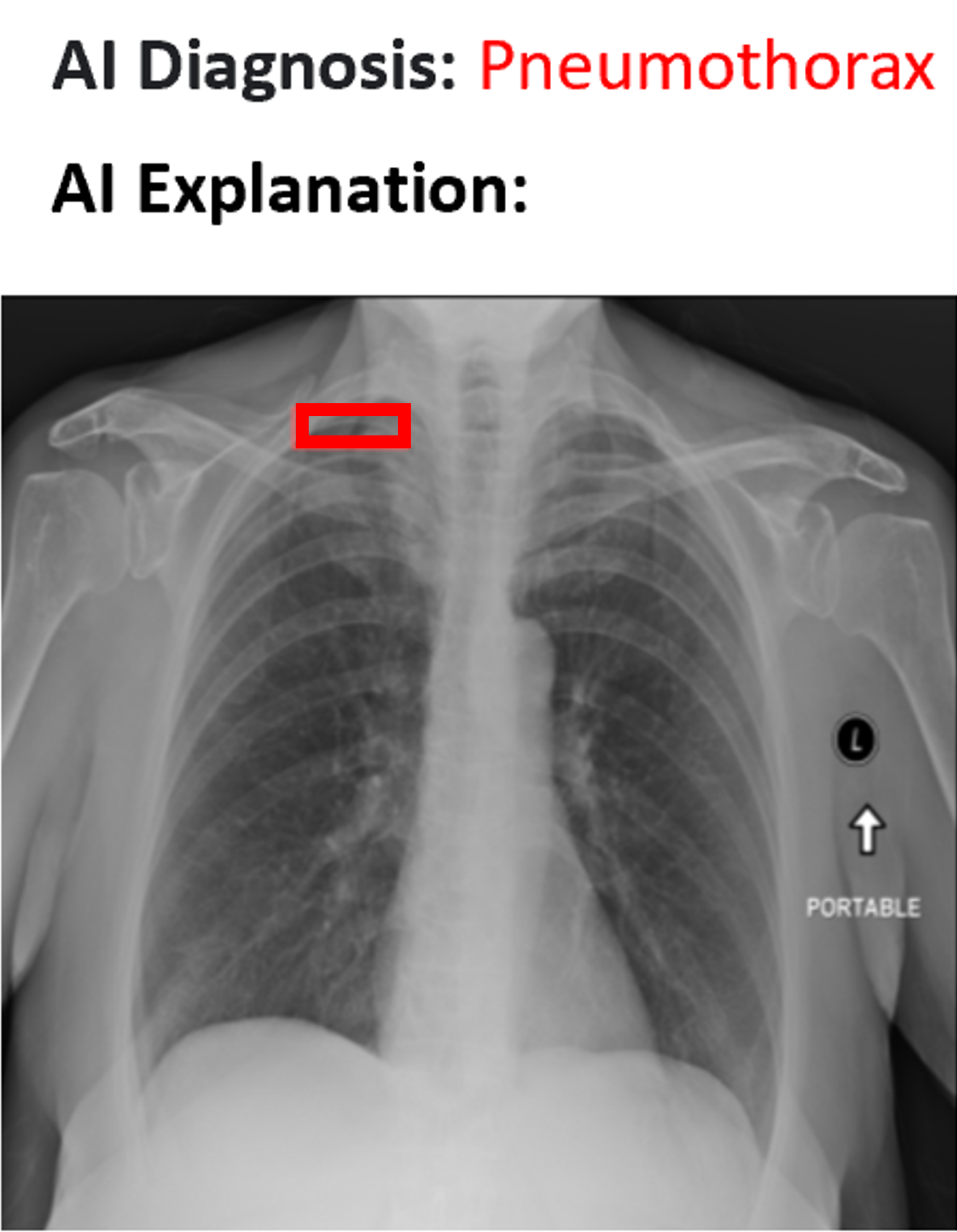}
    \vspace{-0.3cm}
    \caption{Visual XAI: \textit{Bounding Box}.}
    \label{BoundingBox}
\end{subfigure}
    \vspace{-0.1cm}
\par\bigskip 
\begin{subfigure}{0.4\textwidth}
    \includegraphics[width=\linewidth]{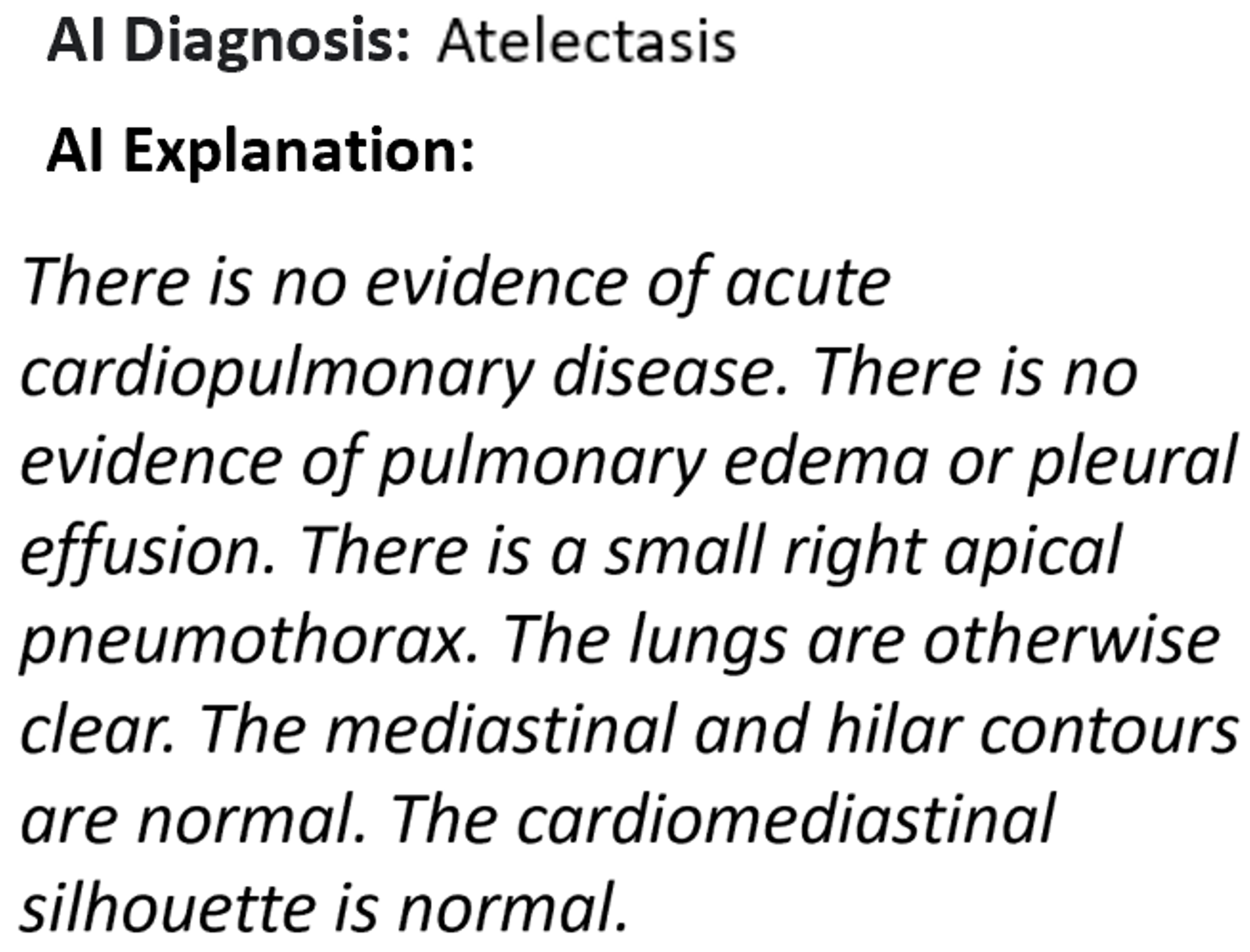}
    \vspace{-0.3cm}
    \caption{Textual XAI: \textit{Report}.}
    \label{Report}
\end{subfigure}\hfill
\begin{subfigure}{0.45\textwidth}
    \includegraphics[width=\linewidth]{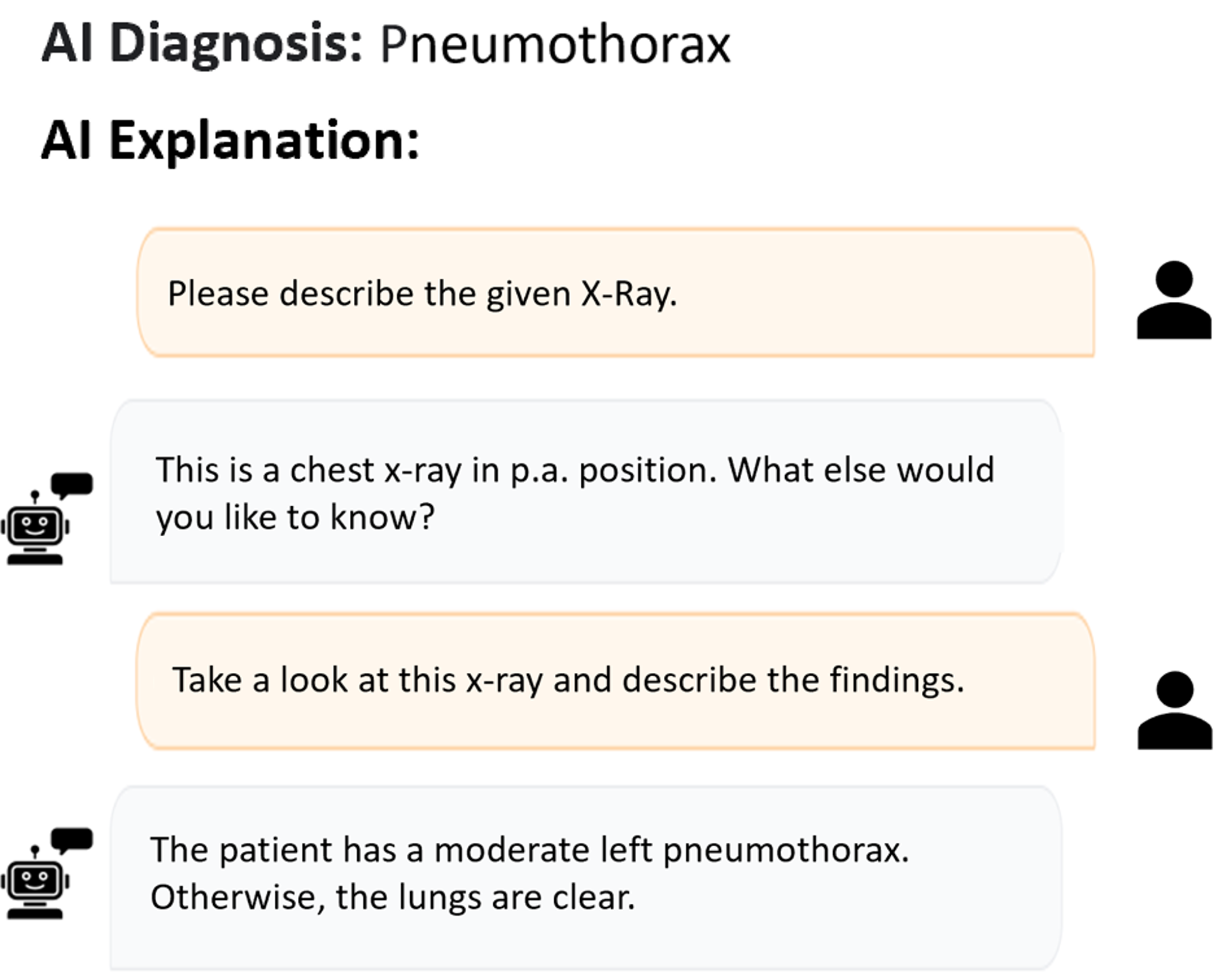}
    \vspace{-0.3cm}
    \caption{Textual XAI: \textit{Chatbot}.}
    \label{Chatbot}
\end{subfigure}
\vspace{-0.1cm}
\caption{Visual and Textual XAI Methods.}
\label{Visual and Textual XAI Methods}
\end{figure}

\section{Analyzed XAI Methods}
\label{XAI Methods in Medical Image Diagnosis}

\vspace{-0.2cm}

This section presents the state-of-the-art visual and textual XAI methods analyzed in our study with physicians. Without these XAI methods, physicians would only have X-ray images.

\vspace{-0.4cm}





\subsection{Visual XAI Methods}
\label{XAI Methods:Visual XAI Methods}


\vspace{-0.2cm}

\subsubsection{Heatmap}

Fig.~\ref{Heatmap} shows a \textit{heatmap} based on the ChestX-ray8 dataset~\cite{Wang:2017} and TorchXrayVision~\cite{pmlr-v172-cohen22a}. Relevant features for neural network's classification are color-coded in the X-ray. 
For example, \cite{Rajpurkar:2017}'s pneumonia detection algorithm uses \textit{heatmaps} with class activation mappings~\cite{zhou2015learning} to identify thoracic diseases~\cite{Wang:2017}.


\vspace{-0.3cm}

\subsubsection{Bounding Box}
\label{subsubsec:Bounding Box}

Fig.~\ref{BoundingBox} shows a \textit{bounding box} added to a ChestX-ray8 image~\cite{Wang:2017}, based on~\cite{Kashyap:2020}, who use GAIN for \textit{bounding boxes}. A \textit{bounding box} indicates disease location, resembling human markings. For example, \cite{Kashyap:2020} derive bounding boxes from reports describing disease locations.

\vspace{-0.5cm}

\subsection{Textual XAI Methods}
\label{XAI Methods:Textual XAI Methods}


\vspace{-0.2cm}

\subsubsection{Report}
\label{subsubsec:Report}

Fig.\ref{Report} shows a medical \textit{report} based on the MIMIC-CXR dataset~\cite{johnson2019mimic}. 
Transformers dominate \textit{report} generation, leveraging models like RadFM~\cite{wu2023generalist} or BioMedGPT~\cite{zhang2024biomedgpt} and datasets such as IU-Xray~\cite{DemnerFushman2015PreparingAC} and MIMIC-CXR~\cite{johnson2019mimic}.


\vspace{-0.4cm}

\subsubsection{Chatbot}
\label{subsubsec:Chatbot}

Fig.~\ref{Chatbot} shows a conversation between a physician and a medical \textit{chatbot} based on the MIMIC-CXR dataset~\cite{johnson2019mimic} and \cite{thawkar2023xraygpt}'s XrayGPT. These \textit{chatbots} provide personalized support to healthcare professionals and patients~\cite{thawkar2023xraygpt}. Large language models power medical \textit{chatbots} such as MedFlamingo~\cite{moor2023medflamingo}, XrayGPT~\cite{thawkar2023xraygpt}, LlaVA-Med~\cite{li2023llavamed}, and ChatCAD+~\cite{zhao2023chatcad}, often fine-tuned with medical data.


\vspace{-0.3cm}

\subsection{Multimodal XAI Methods}
\label{XAI Methods:Multimodal XAI Methods}

\vspace{-0.1cm}


\subsubsection{Heatmap+Report}

Fig.~\ref{Heatmap+Report} combines a \textit{heatmap} and medical \textit{report} using a \textit{bounding box} on a ChestX-ray8 image~\cite{Wang:2017} and a \textit{report} generated like in~\cite{Cheng:2023,Tanida:2023}. Approaches like~\cite{chen2024crossmodal,Cheng:2023} integrate \textit{heatmaps} (e.g., Grad-CAM~\cite{Yuan:2019,Tanwani_2022}) and \textit{reports}. 



\vspace{-0.4cm}

\subsubsection{Heatmap+Chatbot}

Fig.~\ref{Heatmap+Chatbot} integrates a \textit{heatmap} from ChestX-ray8~\cite{Wang:2017} with a \textit{chatbot}-generated description~\cite{pmlr-v172-cohen22a,thawkar2023xraygpt}. MedFuseNet~\cite{Sharma2021}, the only model combining \textit{heatmap} and \textit{chatbot}, uses attention mechanisms for input fusion.

\vspace{-0.4cm}

\subsubsection{Bounding Box+Report}

Fig.~\ref{BoundingBox+Report} demonstrates the combination of \textit{bounding box} and medical \textit{report}, where we added the \textit{report} to an image from the ChestX-ray8 dataset~\cite{Wang:2017} and the corresponding \textit{report} based on~\cite{Cheng:2023,wang2018tienet,Tanida:2023}.  
We identified three implementations of this approach: RGRG~\cite{Tanida:2023},  TieNet~\cite{wang2018tienet} and VLCI~\cite{Cheng:2023}.



\vspace{-0.4cm}

\subsubsection{Bounding Box+Chatbot}

\begin{figure}[h!]
\centering
\begin{subfigure}{0.33\textwidth}
    \includegraphics[width=\linewidth]{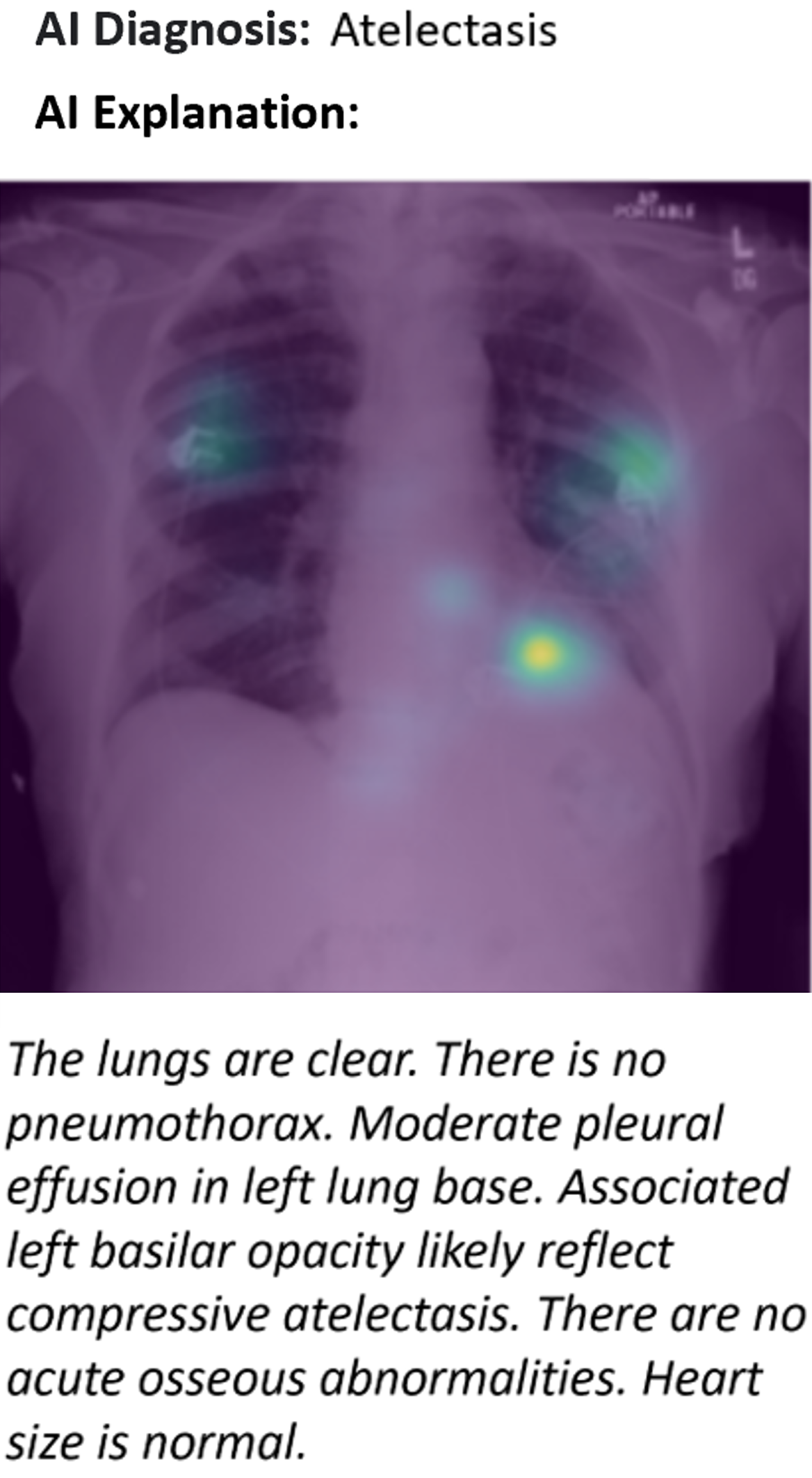}
    \caption{\textit{Heatmap+Report}.}
    \label{Heatmap+Report}
\end{subfigure}\hfill
    \vspace{-0.1cm}
\begin{subfigure}{0.45\textwidth}
    \includegraphics[width=\linewidth]{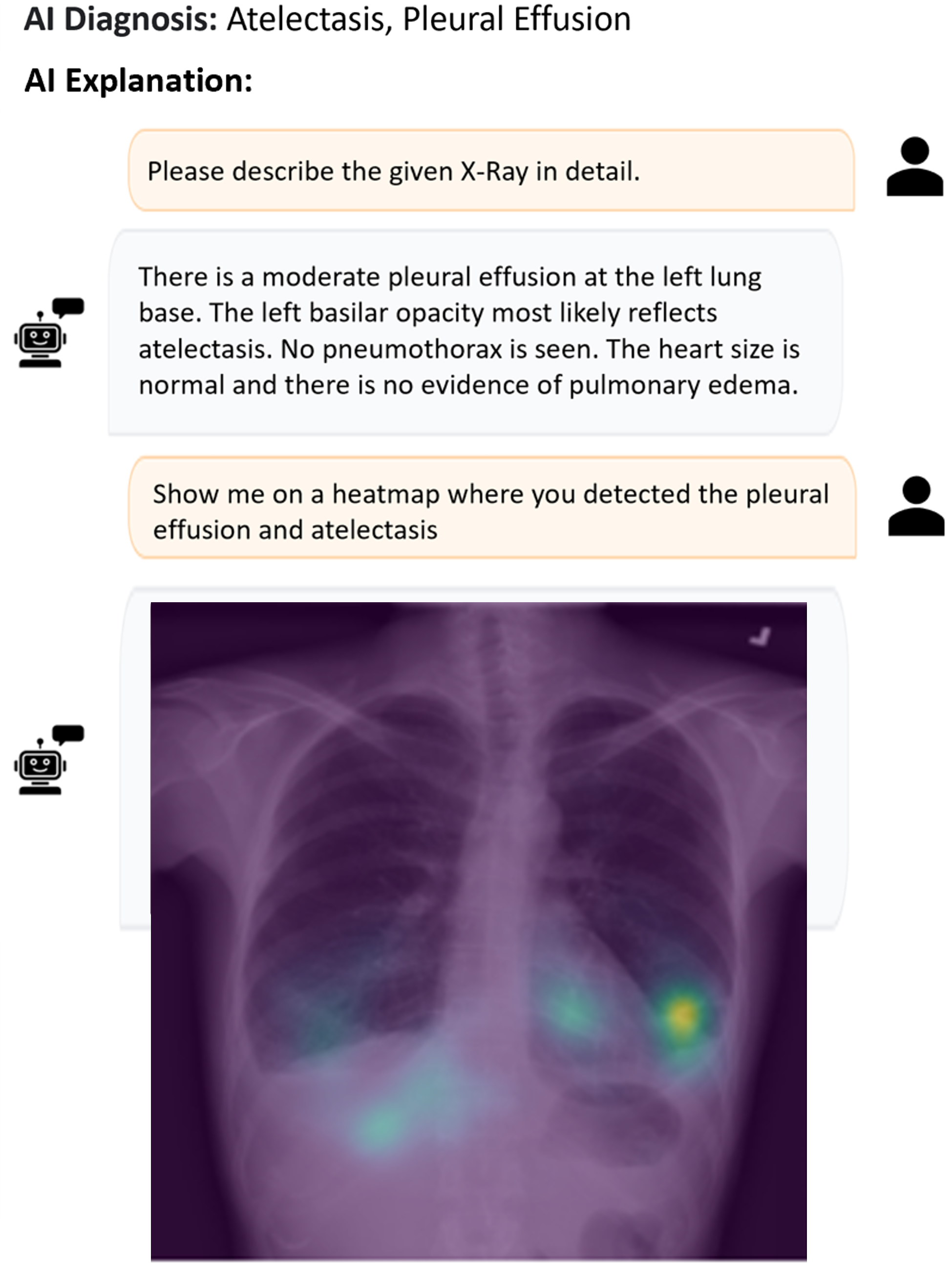}
    \caption{\textit{Heatmap+Chatbot}.}
    \label{Heatmap+Chatbot}
\end{subfigure}
    \vspace{-0.1cm}
\par\bigskip 
\begin{subfigure}{0.38\textwidth}
    \includegraphics[width=\linewidth]{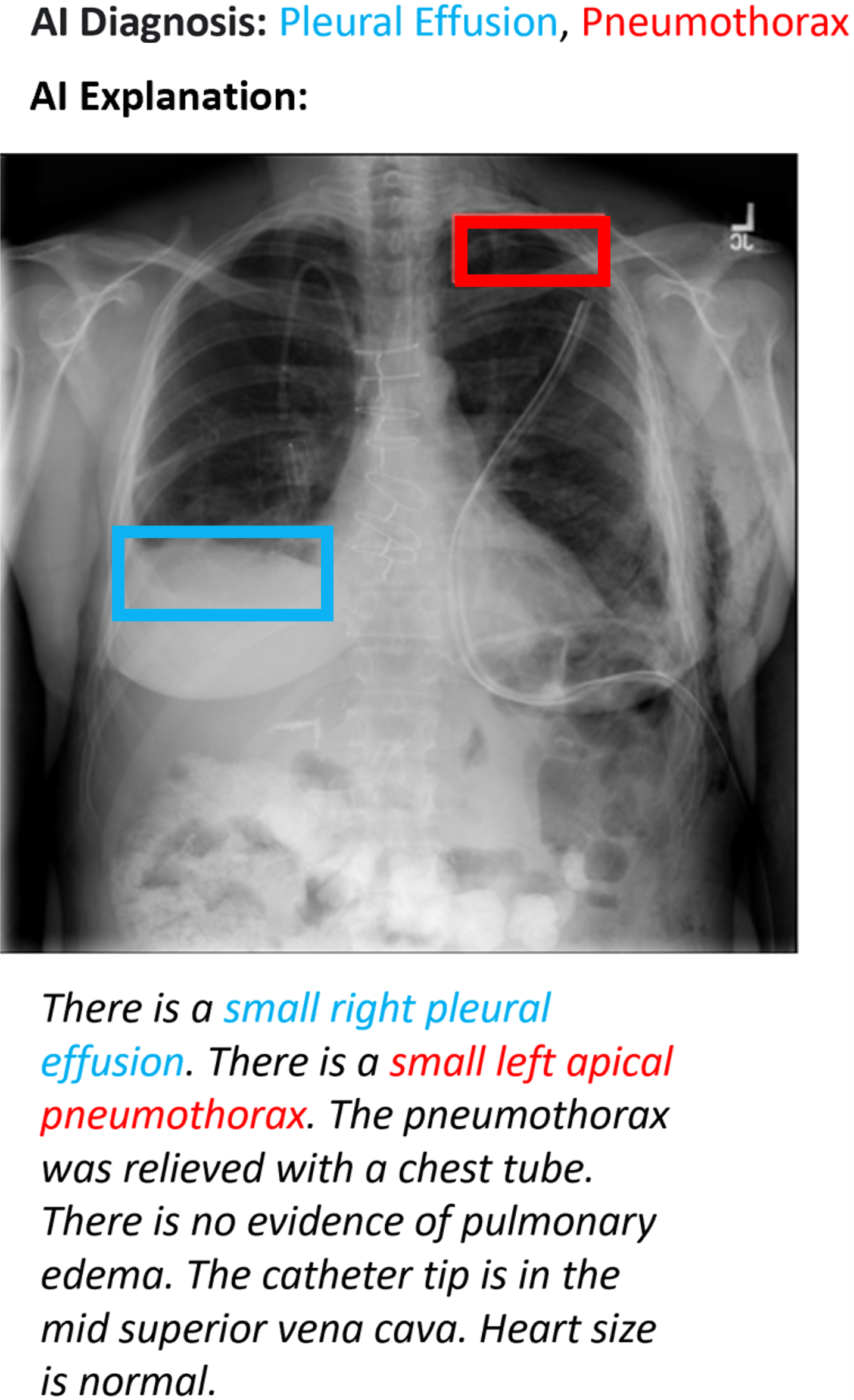}
    \caption{\textit{Bounding Box+Report}.}
    \label{BoundingBox+Report}
\end{subfigure}\hfill
\begin{subfigure}{0.45\textwidth}
    \includegraphics[width=\linewidth]{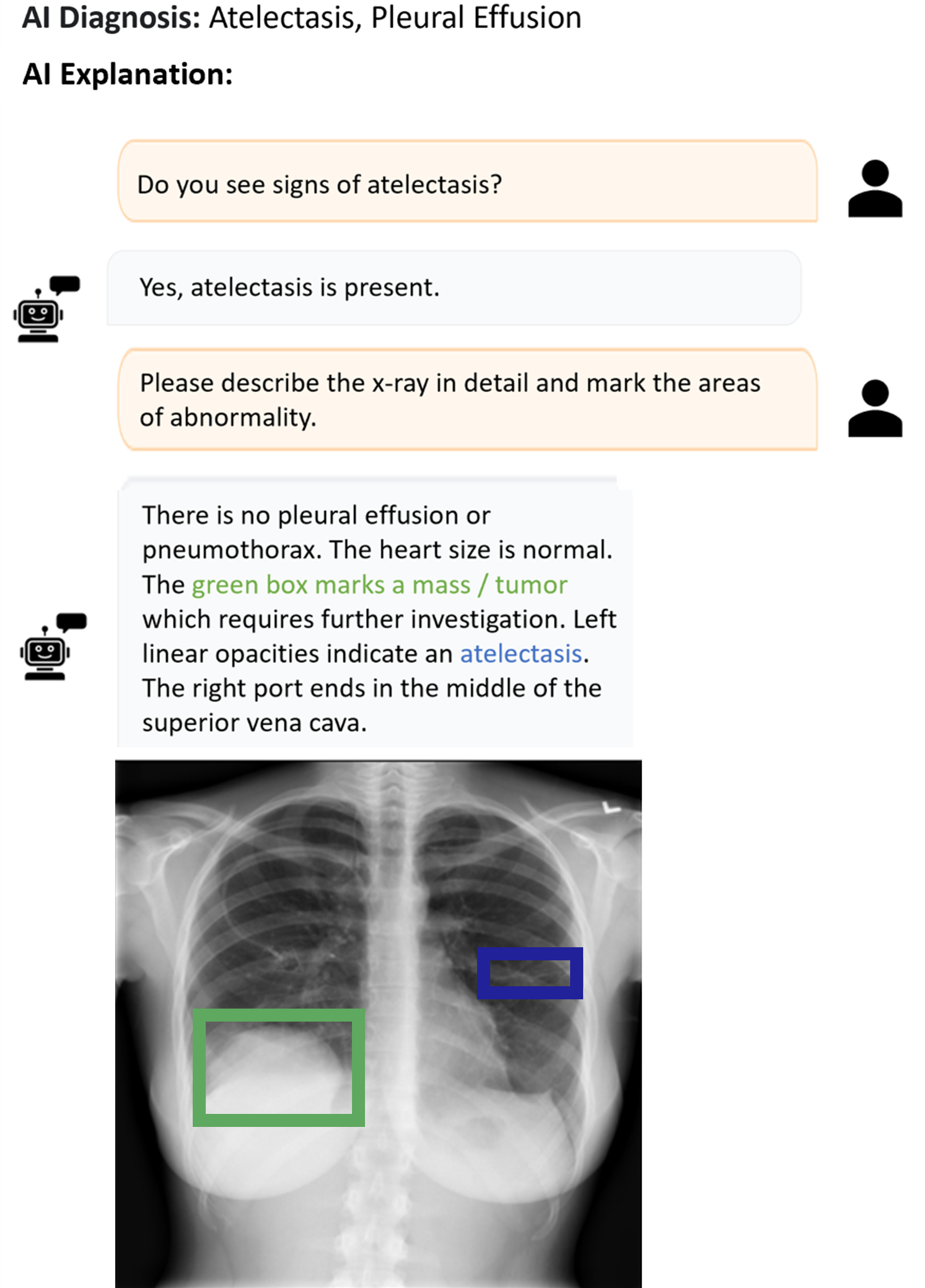}
    \caption{\textit{Bounding Box+Chatbot}.}
    \label{BoundingBox+Chatbot}
\end{subfigure}
\vspace{-0.1cm}
\caption{Multimodal XAI Methods.}
\label{Multimodal XAI Methods}
\end{figure}

Fig.~\ref{BoundingBox+Chatbot} combines a \textit{bounding box} on a ChestX-ray8 image~\cite{Wang:2017} with a \textit{chatbot}-generated description~\cite{thawkar2023xraygpt}. No publication yet describes this combination, but it could be realized by integrating \textit{chatbot} algorithms from Sec.~\ref{subsubsec:Chatbot} with \textit{bounding box} generation methods in Sec.~\ref{subsubsec:Bounding Box}. 


\vspace{-0.8cm}

\section{Questionnaire and Participants}
\label{Experimental Setup}

\vspace{-0.2cm}



Our survey 
evaluated the 8 XAI methods from Sec.~\ref{XAI Methods in Medical Image Diagnosis} plus \textit{no explanation}. 
Participants assessed these methods on \textit{understandability}, \textit{completeness}, \textit{speed}, and \textit{applicability} using a Likert scale (1–5). They also analyzed one correct and one incorrect AI diagnosis per method, with 50\% of diagnoses being incorrect. 
The questionnaire was completed by 33~participants (24~female, 8~male, 1~diverse): 49\% assistant physicians, 24\% specialists, 12\% senior physicians, and 15\% medical students. 
34\% had 1–5 years, 18\% had 5–10 years, 18\% had 10–20 years, 12\% had over 20 years, and 18\% had no experience. Specializations included general practitioners/internal specialists (28\%), radiology, anesthesiology, surgery, and orthopedics (12\% each), pediatrics/psychiatry (6\%), and 18\% undecided. 
Most participants (76\%) \textit{strongly agree} they were open to new technologies, 15\% \textit{agree}. However, 45\% \textit{strongly disagree} and 18\% \textit{disagree} about having experience with AI-based diagnosis, showing limited practice use. Positive AI experiences were reported by 27\% (18\% \textit{agree}, 9\% \textit{strongly agree}). 
The remaining 10\% were \textit{neutral}, reflecting uncertainty about which systems use AI.


\vspace{-0.4cm}

\begin{figure}[h!]
\centering
\begin{subfigure}{0.49\textwidth}
    \includegraphics[width=\linewidth]{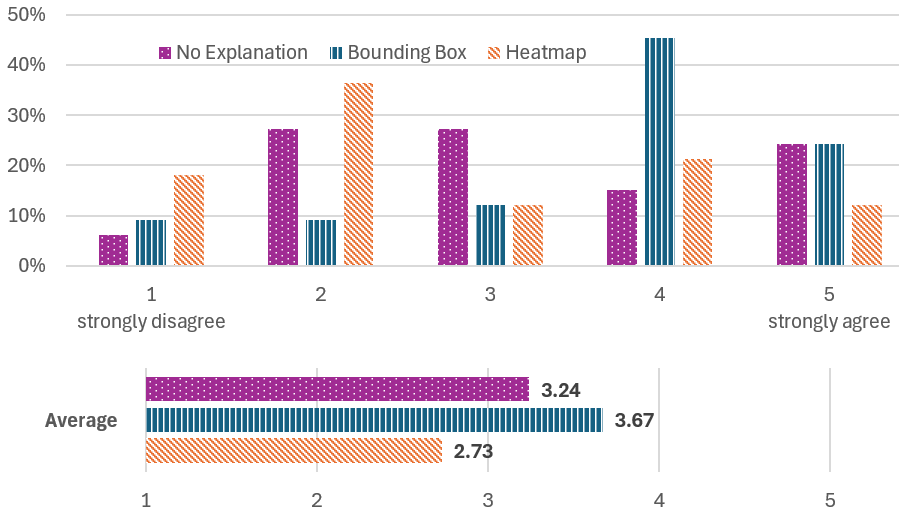}
    \vspace{-0.3cm}
    \caption{\textit{Understandability}.}
    \label{VisualXAI_Understandability}
\end{subfigure}\hfill
    \vspace{-0.1cm}
\begin{subfigure}{0.49\textwidth}
    \includegraphics[width=\linewidth]{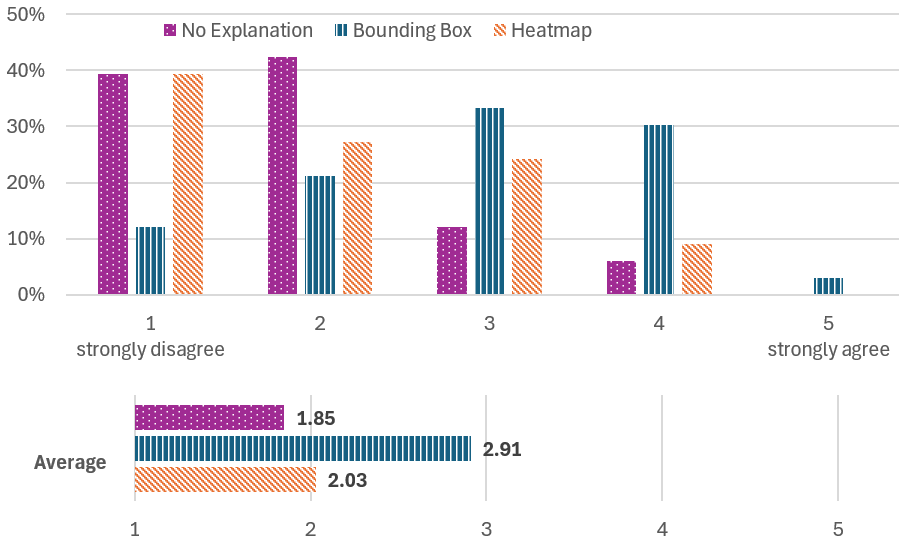}
    \vspace{-0.3cm}
    \caption{\textit{Completeness}.}
    \label{VisualXAI_Completeness}
\end{subfigure}
\par\bigskip 
    \vspace{-0.1cm}
\begin{subfigure}{0.49\textwidth}
    \includegraphics[width=\linewidth]{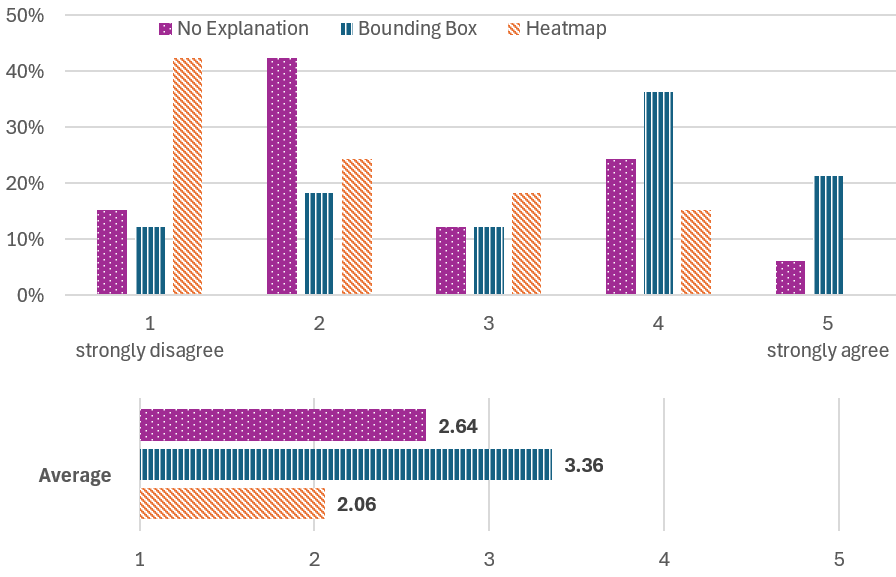}
    \vspace{-0.3cm}
    \caption{\textit{Speed}.}
    \label{VisualXAI_Speed}
\end{subfigure}\hfill
\begin{subfigure}{0.49\textwidth}
    \includegraphics[width=\linewidth]{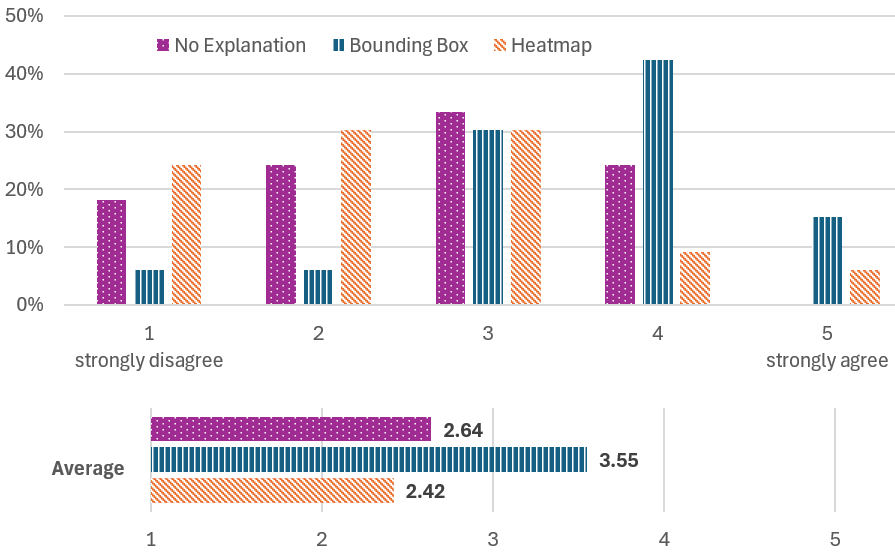}
    \vspace{-0.3cm}
    \caption{\textit{Applicability}.}
    \label{VisualXAI_Applicability}
\end{subfigure}
\vspace{-0.2cm}
\caption{Visual XAI Methods.}
\label{Visual XAI Methods}
\end{figure}

\vspace{-0.8cm}

\section{Experiments and Results}
\label{Experiments and Results}

\vspace{-0.2cm}


\subsection{Visual XAI Methods}
\label{Experiments and Results:Visual XAI Methods}

\vspace{-0.1cm}


\subsubsection{Understandability}
\label{Experiments and Results:Visual XAI Methods:Understandability}

Fig.~\ref{VisualXAI_Understandability} shows the assessment of the \textit{understandability} of visual XAI methods. \textit{bounding box} scores highest with an average of 3.67, 18.68\% higher than the 2nd-best \textit{no explanation}~(3.24). Most participants (45\%) \textit{agree} and 24\% \textit{strongly agree} on its \textit{understandability}. \textit{heatmap} scores lowest at 2.73, indicating difficulty in understandability. Our t-test shows a significant difference between \textit{bounding box} and \textit{heatmap}.


\vspace{-0.5cm}

\subsubsection{Completeness}
\label{Experiments and Results:Visual XAI Methods:Completeness}

Fig.~\ref{VisualXAI_Completeness} shows that almost no participant \textit{strongly agrees} on the \textit{completeness} of visual XAI methods or \textit{no explanation}. Only 3\% \textit{strongly agree} that \textit{bounding box} is sufficiently complete, scoring highest with 2.91, 43.35\% higher than \textit{heatmap}. Our t-test confirms a significant difference. 
\textit{No explanation} scores lowest at 1.85, with 39\% \textit{strongly disagreeing} and 42\% \textit{disagreeing}. All visual XAI methods rated below average, indicating insufficient information for understanding AI diagnoses.



\vspace{-0.5cm}

\subsubsection{Speed}
\label{Experiments and Results:Visual XAI Methods:Speed}

Fig.~\ref{VisualXAI_Speed} illustrates participant evaluations of visual XAI methods for \textit{speed}. \textit{bounding box} was preferred, averaging 3.36, 27.27\% higher than \textit{no explanation} (2.64). While 42\% \textit{disagree} that \textit{no explanation} speeds up diagnosis, 24\% \textit{agree}. \textit{heatmap} scores lowest at 2.06, with 42\% \textit{strongly disagreeing} it aids faster diagnosis. A significant difference 
between \textit{no explanation} and \textit{heatmap} shows that \textit{heatmap} slows down diagnosis.

\vspace{-0.5cm}

\subsubsection{Applicability}
\label{Experiments and Results:Visual XAI Methods:Applicability}

Fig.~\ref{VisualXAI_Applicability} presents participant assessments of visual XAI \textit{applicability}. \textit{bounding box} leads with a 3.55 average score, with 42\% \textit{strongly agreeing} on its practical use, while only 12\% \textit{disagree}. \textit{no explanation} averages 2.64, and \textit{heatmap} scores lowest at 2.42. For \textit{heatmap}, 54\% of participants \textit{strongly disagree}, \textit{disagree}, or were \textit{neutral} about its applicability. Only 6\% \textit{strongly agree} that \textit{heatmap} could be used in practice, compared to none for \textit{no explanation}.



\vspace{-0.4cm}

\begin{figure}[h!]
\centering
\begin{subfigure}{0.49\textwidth}
    \includegraphics[width=\linewidth]{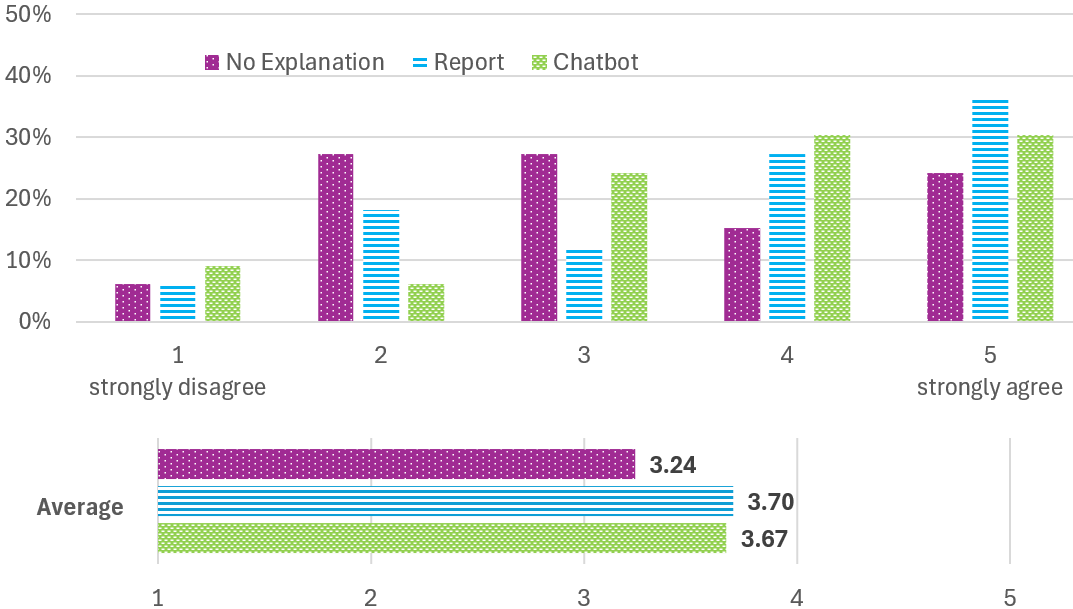}
    \vspace{-0.4cm}
    \caption{\textit{Understandability}.}
    \label{TextualXAI_Understandability}
\end{subfigure}\hfill
\begin{subfigure}{0.49\textwidth}
    \includegraphics[width=\linewidth]{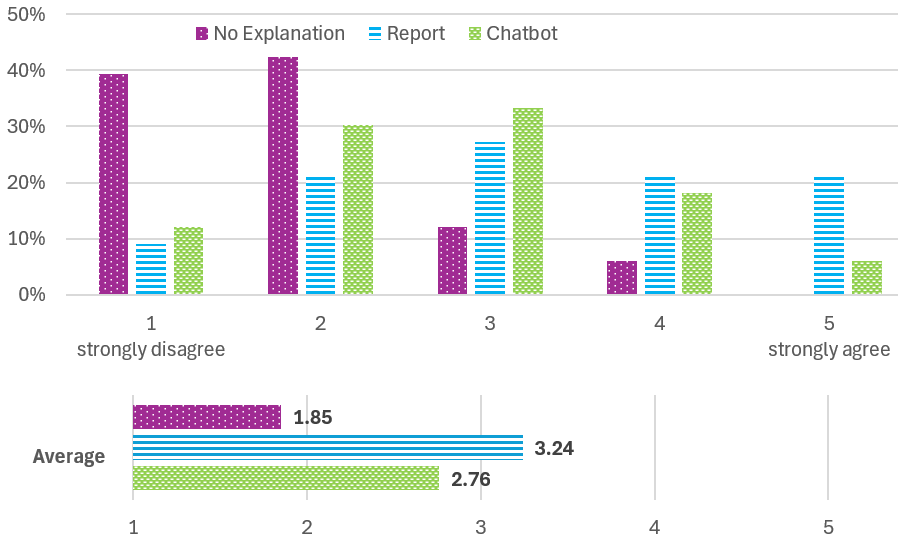}
    \vspace{-0.4cm}
    \caption{\textit{Completeness}.}
    \label{TextualXAI_Completeness}
\end{subfigure}
\par\bigskip 
\begin{subfigure}{0.49\textwidth}
    \includegraphics[width=\linewidth]{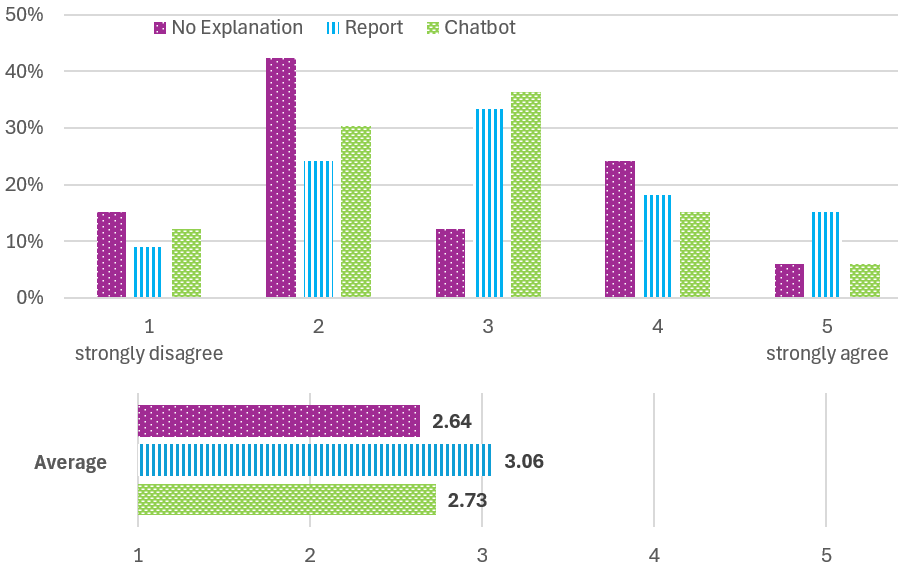}
    \vspace{-0.4cm}
    \caption{\textit{Speed}.}
    \label{TextualXAI_Speed}
\end{subfigure}\hfill
\begin{subfigure}{0.49\textwidth}
    \includegraphics[width=\linewidth]{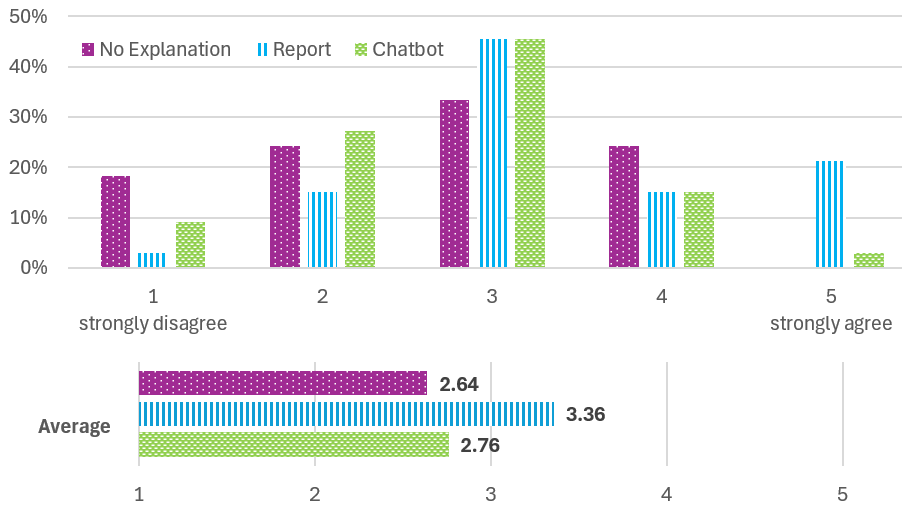}
    \vspace{-0.4cm}
    \caption{\textit{Applicability}.}
    \label{TextualXAI_Applicability}
\end{subfigure}
\vspace{-0.1cm}
\caption{Textual XAI Methods.}
\label{Textual XAI Methods}
\end{figure}


\subsection{Textual XAI Methods}


\vspace{-0.1cm}

\subsubsection{Understandability}

Fig.~\ref{TextualXAI_Understandability} shows the evaluations of textual XAI methods for \textit{understandability}. \textit{report} scores 3.70, slightly above \textit{chatbot} (3.67) and higher than \textit{no explanation} (3.24). Significant differences exist between \textit{report} 
/ \textit{chatbot} 
and \textit{no explanation}. Most participants \textit{agree} (27\%) or \textit{strongly agree} (36\%) that \textit{report} is understandable, with similar results for \textit{chatbot} (30\% agree, 30\% strongly agree). But 6\%-9\% \textit{strongly disagree} with understanding all methods.


\vspace{-0.4cm}

\subsubsection{Completeness}

Fig.~\ref{TextualXAI_Completeness} shows the evaluations of the textual XAI methods' \textit{completeness}. \textit{report} ranks highest (3.24), followed by \textit{chatbot} (2.76), both significantly above \textit{no explanation} (1.85). Statistically significant differences exist between \textit{chatbot} vs. \textit{no explanation} 
and \textit{report} vs. \textit{no explanation}. 
For \textit{report}, opinions are divided: 21\% \textit{disagree}, 27\% \textit{neutral}, 21\% \textit{agree}, and 21\% \textit{strongly agree} that \textit{report} contains all necessary information to understand the AI diagnosis. Only 9\% \textit{strongly disagree}. For \textit{chatbot}, 33\% \textit{neutral} and 30\% \textit{disagree}. Compared to Sec.~\ref{Experiments and Results:Visual XAI Methods:Completeness}, \textit{report} also outperforms the visual XAI methods \textit{bounding box} (2.91) and \textit{heatmap} (2.03) in terms of \textit{completeness}. Familiarity with the \textit{report} format, commonly used by radiologists, may explain this.


\vspace{-0.4cm}

\subsubsection{Speed}

Fig.~\ref{TextualXAI_Speed} shows the evaluations of textual XAI methods' \textit{speed}. \textit{report} leads with 3.06 on average, outperforming \textit{chatbot} (2.73) by 12.09\%, both above \textit{no explanation} (2.64), but close to average, indicating minimal effect on diagnosis speed. \textit{Report} appears neutral, while \textit{chatbot} may slightly delay due to longer text. Compared to visual XAI methods like \textit{bounding box} (3.36), textual methods may be slower, especially when visual aids like \textit{heatmap} (2.06) are less clear.
 

\vspace{-0.4cm}

\subsubsection{Applicability}

Fig.~\ref{TextualXAI_Applicability} presents the assessments of the textual XAI methods' \textit{applicability}. \textit{report} scores highest (3.36), with 21\% \textit{strongly agreeing} on its practical use, but 45\% rated textual methods as \textit{neutral}. \textit{chatbot} averages 2.76, outperforming \textit{no explanation} (2.64) by 27\%. Our t-test shows a significant difference between \textit{report} and \textit{no explanation}, 
but not between \textit{chatbot} and \textit{no explanation}. 
Participants rate textual XAI higher than \textit{no explanation}, but remain skeptical about \textit{chatbot} and more positive about \textit{report}. 



\begin{figure}[h!]
\centering
\begin{subfigure}{0.49\textwidth}
    \includegraphics[width=\linewidth]{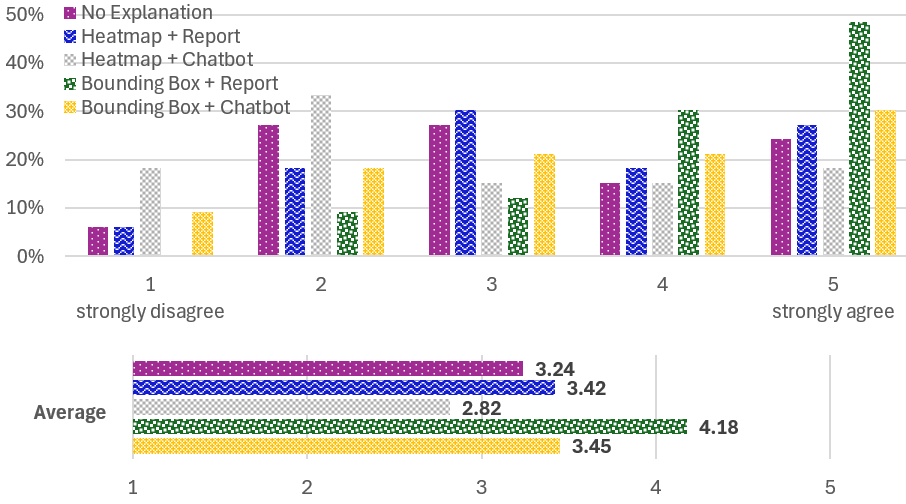}
    \vspace{-0.3cm}
    \caption{\textit{Understandability}.}
    \label{Visual+TextualXAI_Understandability}
\end{subfigure}\hfill
\begin{subfigure}{0.49\textwidth}
    \includegraphics[width=\linewidth]{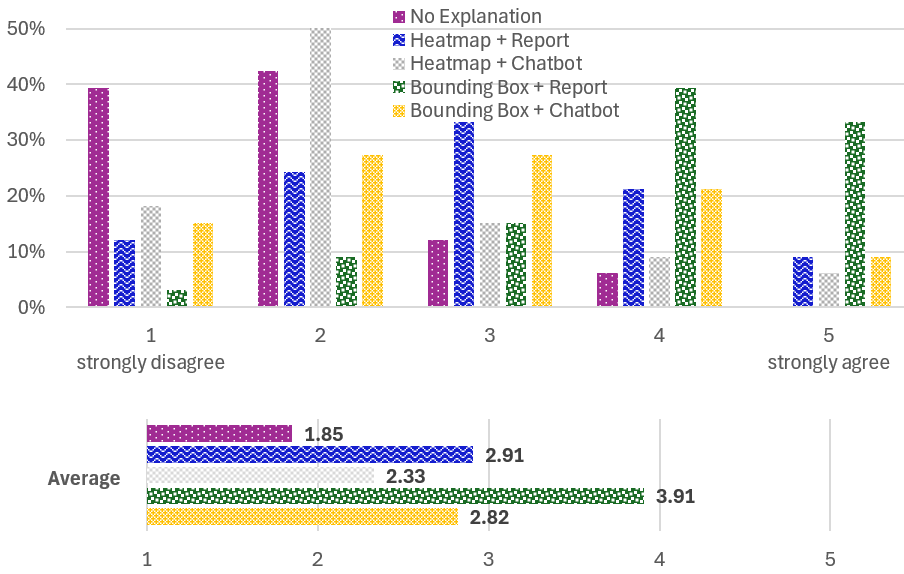}
    \vspace{-0.3cm}
    \caption{\textit{Completeness}.}
    \label{Visual+TextualXAI_Completeness}
\end{subfigure}
\par\bigskip 
\begin{subfigure}{0.49\textwidth}
    \includegraphics[width=\linewidth]{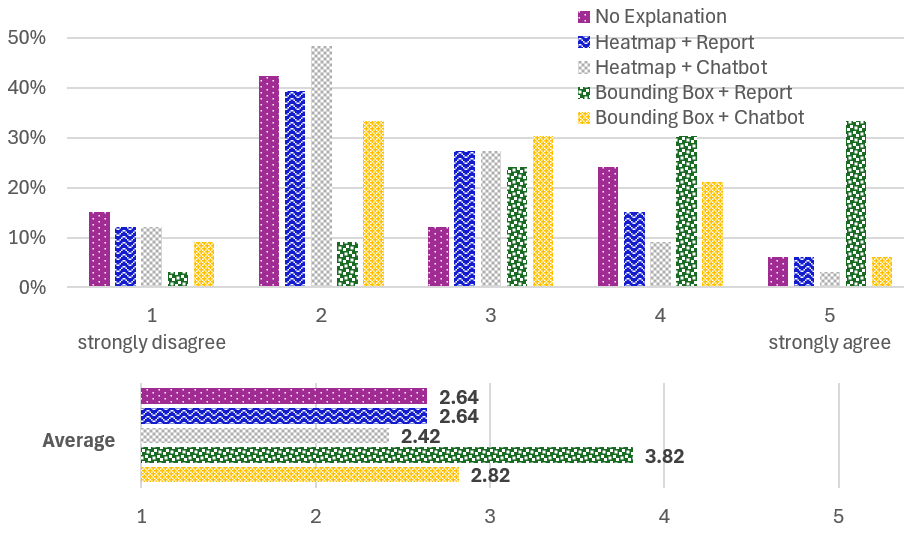}
    \vspace{-0.3cm}
    \caption{\textit{Speed}.}
    \label{Visual+TextualXAI_Speed}
\end{subfigure}\hfill
\begin{subfigure}{0.49\textwidth}
    \includegraphics[width=\linewidth]{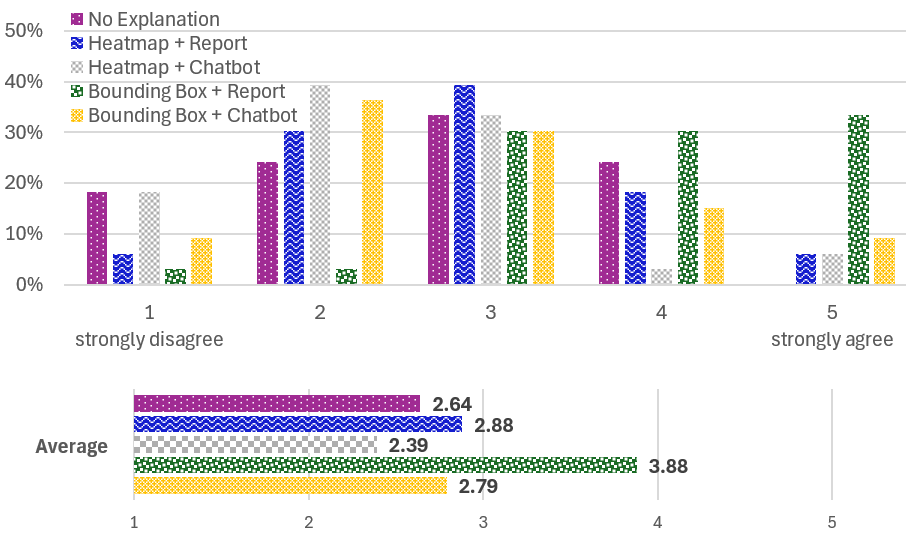}
    \vspace{-0.3cm}
    \caption{\textit{Applicability}.}
    \label{Visual+TextualXAI_Applicability}
\end{subfigure}
\vspace{-0.1cm}
\caption{Multimodal XAI Methods.}
\label{Multimodal XAI Methods}
\end{figure}

\vspace{-0.4cm}

\subsection{Multimodal XAI Methods}


\vspace{-0.2cm}

\subsubsection{Understandability}

In Fig.~\ref{Visual+TextualXAI_Understandability}, participants rate \textit{bounding box}+\textit{report} as the most understandable multimodal XAI method (4.18): 48\% \textit{strongly agree} and 30\% \textit{agree}. It outperforms \textit{bounding box}+\textit{chatbot} (3.45) by 21\% and \textit{report} (3.70) alone by 13\%. Our t-test confirms statistical significance. 
This highlights the value of analyzing multimodal combinations separately from unimodal methods. 



\vspace{-0.4cm}

\subsubsection{Completeness}

Fig.~\ref{Visual+TextualXAI_Completeness} shows the evaluations of the multimodal XAI methods' \textit{completeness}. \textit{bounding box}+\textit{report} leads with 3.91, with 33\% \textit{strongly agreeing} and 30\% \textit{agreeing}. The 2nd-best, \textit{report}+\textit{heatmap}, scores 2.91, followed by \textit{bounding box}+\textit{chatbot} at 2.82. The lowest-rated, \textit{heatmap}+\textit{chatbot}, scores 2.33, yet still outperforms \textit{no explanation} by 26\%. 

\vspace{-0.4cm}

\subsubsection{Speed}

Fig.~\ref{Visual+TextualXAI_Speed} shows the evaluations on the multimodal XAI methods' \textit{speed}. Most \textit{disagree} that the methods speed up diagnosis: 33\% for \textit{bounding box}+\textit{chatbot}, 39\% for \textit{heatmap}+\textit{report}, 42\% for \textit{no explanation}, and 48\% for \textit{heatmap}+\textit{chatbot}, with average scores from 2.42 to 2.82. But 33\% \textit{strongly agree} that \textit{bounding box}+\textit{report} improves \textit{speed}, scoring 3.82, making it the top performer compared to both visual and textual XAI methods. The only other methods scoring above the average of~3 are the unimodal versions: \textit{bounding box} (3.36) and \textit{report} (3.06). 


\vspace{-0.4cm}

\subsubsection{Applicability}

Fig.~\ref{Visual+TextualXAI_Applicability} shows participants' assessment of the \textit{applicability} of multimodal XAI methods. \textit{bounding box}+\textit{report} leads with an average score of 3.88. 33\% \textit{strongly agree} on its practical use. It significantly outperforms \textit{heatmap}+\textit{report} (2.88), with t(32)=5.0143; p=0.0001 $<$ 0.05. \textit{bounding box}+\textit{chatbot}, scores 2.64, while \textit{heatmap}+\textit{chatbot} (2.39) scores lower than \textit{no explanation} (2.64). 
\textit{bounding box}+\textit{report} is the most applicable multimodal XAI method. For comparison: The only unimodal methods exceeding an average score of 3 are its components: \textit{bounding box} (3.55) and \textit{report} (3.36). 




\vspace{-0.4cm}

\subsection{Influence on Desicion-Making}

\vspace{-0.1cm}

\begin{figure}[h!]
\centering
\begin{subfigure}{0.49\textwidth}
    \includegraphics[width=\linewidth]{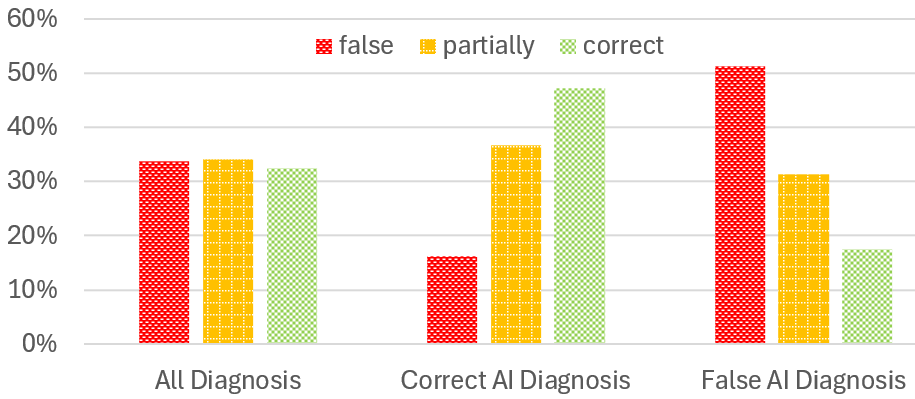}
    \vspace{-0.5cm}
    \caption{Total Distribution.}
\label{FalseDiagnoses_all}
\end{subfigure}\hfill
    \vspace{-0.5cm}
\begin{subfigure}{0.49\textwidth}
    \includegraphics[width=\linewidth]{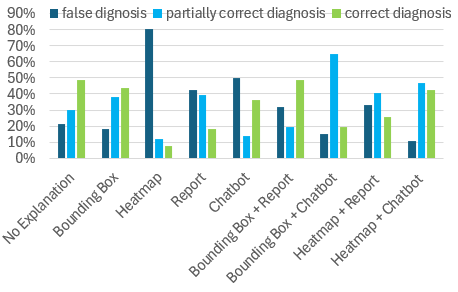}
    \vspace{-0.1cm}
    \caption{Distribution over XAI Methods.}
    \label{FalseDiagnoses}
\end{subfigure}
\par\bigskip 
\vspace{-0.2cm}
\caption{Correct, Partly Correct and False Diagnoses.}
\label{Correct, Partly Correct and False Diagnoses}
\end{figure}

Finally, we analyzed the \textit{influence} of XAI methods on physicians' decisions with correct and incorrect AI predictions. They viewed 1~correct and 1~incorrect diagnosis per method. Fig.~\ref{FalseDiagnoses_all} shows 50\% trust false AI diagnoses across all methods. Fig.~\ref{FalseDiagnoses} highlights \textit{heatmap} as causing the most \textit{false diagnoses}, while \textit{bounding box}+\textit{chatbot} yields the most \textit{partially correct} diagnoses. Even \textit{bounding box}+\textit{report}, the top method, results in 32\% \textit{false} and 20\% \textit{partially false} diagnoses, stressing the need for fewer incorrect AI predictions to maintain trust. 


\vspace{-0.5cm}

\vspace{-1.1cm}



\section{Conclusion and Future Work}
\label{Conclusion and Future Work}

\vspace{-0.3cm}

Our findings emphasize physicians' demand for AI explanations and the need to integrate AI into clinical practice. \textit{bounding box}+\textit{report} emerged as the most effective and comprehensive XAI method. Concerns about false AI diagnoses underline the importance of accurate systems and physician training. 

While this study focused on visual and textual methods, future work could explore example-based XAI and integrate patient history into models, requiring large datasets with annotated x-rays and histories.

\vspace{-0.2cm}

\begin{credits}
\subsubsection{\ackname} This research was funded by IU International University of Applied Sciences (\textit{IU Incubator}) from October 2023 to September 2025.

\vspace{-0.35cm}

\subsubsection{\discintname}
The authors have no competing interests to declare.
\end{credits}

\vspace{-0.3cm}

%
%
%
\bibliographystyle{splncs04}
\bibliography{mybibliography}

@String(CVPR= {IEEE Conf. Comput. Vis. Pattern Recog.})

@String(ICCV= {Int. Conf. Comput. Vis.})

@String(CVPR  = {CVPR})

@String(ICCV  = {ICCV})

@article{Rajpurkar:2017,
  author       = {Pranav Rajpurkar and
                  others},
  title        = {{CheXNet: Radiologist-Level Pneumonia Detection on Chest X-Rays with
                  Deep Learning}},
  journal      = {CoRR},
  year         = {2017},
  eprinttype    = {arXiv}

}

@ARTICLE{Adadi:2018,
  author={Adadi, Amina and Berrada, Mohammed},
  journal={IEEE Access}, 
  title={{Peeking Inside the Black-Box: A Survey on Explainable Artificial Intelligence (XAI)}}, 
  year={2018},
  volume={6},
  number={},
  pages={52138-52160}}

@article{VANDERVELDEN2022102470,
title = {{Explainable Artificial Intelligence (XAI) in Deep Learning-Based Medical Image Analysis}},
journal = {Medical Image Analysis},
volume = {79},
pages = {102470},
year = {2022},
issn = {1361-8415},
author = {Bas H.M. {van der Velden} and Hugo J. Kuijf and Kenneth G.A. Gilhuijs and Max A. Viergever}
}

@Article{app122211750,
AUTHOR = {Ahmed, Saad Bin and Solis-Oba, Roberto and Ilie, Lucian},
TITLE = {{Explainable-AI in Automated Medical Report Generation Using Chest X-ray Images}},
JOURNAL = {Applied Sciences},
VOLUME = {12},
YEAR = {2022},
NUMBER = {22},
ARTICLE-NUMBER = {11750},
ISSN = {2076-3417}
}

@misc{prentzas2023explainable,
      title={{Explainable AI Applications in the Medical Domain: A Systematic Review}}, 
      author={Nicoletta Prentzas and Antonis Kakas and Constantinos S. Pattichis},
      year={2023},
      eprint={2308.05411},
      archivePrefix={arXiv},
      primaryClass={cs.AI}
}

@Article{s23020634,
AUTHOR = {Chaddad, Ahmad and Peng, Jihao and Xu, Jian and Bouridane, Ahmed},
TITLE = {{Survey of Explainable AI Techniques in Healthcare}},
JOURNAL = {Sensors},
VOLUME = {23},
YEAR = {2023},
NUMBER = {2},
ARTICLE-NUMBER = {634},
PubMedID = {36679430},
ISSN = {1424-8220},
}

@article{borys2023explainable,
  title={{Explainable AI in Medical Imaging: An Overview for Clinical Practitioners - Beyond Saliency-Based XAI Approaches}},
  author={Borys, Katarzyna and others},
  journal={Eur J Radiol},
  volume={162},
  year={2023}
}

@misc{zhou2015learning,
      title={{Learning Deep Features for Discriminative Localization}}, 
      author={Bolei Zhou and Aditya Khosla and Agata Lapedriza and Aude Oliva and Antonio Torralba},
      year={2015},
      eprint={1512.04150},
      archivePrefix={arXiv},
      primaryClass={cs.CV}
}

@INPROCEEDINGS{Wang:2017,
  author={Wang, Xiaosong and Peng, Yifan and Lu, Le and Lu, Zhiyong and Bagheri, Mohammadhadi and Summers, Ronald M.},
  booktitle={CVPR}, 
  title={{ChestX-Ray8: Hospital-Scale Chest X-Ray Database and Benchmarks on Weakly-Supervised Classification and Localization of Common Thorax Diseases}}, 
  year={2017},
  volume={},
  number={}}

@misc{WHO2009,
  organization = {{World Health Organization (WHO)}},
  title = {{Health Workforce: The Health Workforce Crisis}},
  year = {2009},
  howpublished = {https://www.who.int/news-room/questions-and-answers/item/q-a-on-the-health-workforce-crisis},
  urldate = {2009-09-24},
  note = {Accessed: 2024-03-23}
}

@INPROCEEDINGS{Kashyap:2020,
  author={Kashyap, Satyananda and others},
  booktitle={IEEE ISBI}, 
  title={{Looking in the Right Place for Anomalies: Explainable AI Through Automatic Location Learning}}, 
  year={2020},
  volume={},
  number={},
  pages={1125-1129}}

@misc{zhang2024biomedgpt,
      title={{BiomedGPT: A Unified and Generalist Biomedical Generative Pre-trained Transformer for Vision, Language, and Multimodal Tasks}}, 
      author={Kai Zhang and others},
      year={2024},
      eprint={2305.17100},
      archivePrefix={arXiv},
      primaryClass={cs.CL}
}

@misc{wu2023generalist,
      title={{Towards Generalist Foundation Model for Radiology by Leveraging Web-scale 2D\&3D Medical Data}}, 
      author={Chaoyi Wu and Xiaoman Zhang and Ya Zhang and Yanfeng Wang and Weidi Xie},
      year={2023},
      eprint={2308.02463},
      archivePrefix={arXiv},
      primaryClass={cs.CV}
}

@article{DemnerFushman2015PreparingAC,
  title={{Preparing a Collection of Radiology Examinations for Distribution and Retrieval}},
  author={Dina Demner-Fushman and others},
  journal={Journal of the American Medical Informatics Association},
  year={2015},
  volume={23 2},
  pages={304-10},
}

@article{johnson2019mimic,
  title={{MIMIC-CXR, a De-identified Publicly Available Database of Chest Radiographs with Free-text Reports}},
  author={Johnson, A. E. W. and others},
  journal={Scientific Data},
  volume={6},
  number={1},
  pages={317},
  year={2019}
}

@misc{thawkar2023xraygpt,
      title={{XrayGPT: Chest Radiographs Summarization using Medical Vision-Language Models}}, 
      author={Omkar Thawkar and others},
      year={2023},
      eprint={2306.07971},
      archivePrefix={arXiv},
      primaryClass={cs.CV}
}

@misc{moor2023medflamingo,
      title={{Med-Flamingo: a Multimodal Medical Few-shot Learner}}, 
      author={Michael Moor and others},
      year={2023},
      eprint={2307.15189},
      archivePrefix={arXiv},
      primaryClass={cs.CV}  
}

@misc{li2023llavamed,
      title={{LLaVA-Med: Training a Large Language-and-Vision Assistant for Biomedicine in One Day}}, 
      author={Chunyuan Li and others},
      year={2023},
      eprint={2306.00890},
      archivePrefix={arXiv},
      primaryClass={cs.CV}
}

@misc{zhao2023chatcad,
      title={{ChatCAD+: Towards a Universal and Reliable Interactive CAD using LLMs}}, 
      author={Zihao Zhao and others},
      year={2023},
      eprint={2305.15964},
      archivePrefix={arXiv},
      primaryClass={cs.CV}
}

@InProceedings{pmlr-v172-cohen22a,
  title = 	 {{TorchXRayVision: A Library of Chest X-ray Datasets and Models}},
  author =       {Cohen, Joseph Paul and others},
  booktitle = 	 {Int. Conference on Medical Imaging with Deep Learning},
  pages = 	 {231--249},
  year = 	 {2022},
  volume = 	 {172}
}

@inbook{Tanwani_2022,
   title={{RepsNet: Combining Vision with Language for Automated Medical Reports}},
   booktitle= {MICCAI},
    author={Tanwani, Ajay K. and Barral, Joelle and Freedman, Daniel},
   year={2022},
   pages={714–724} }

@article{Sharma2021,
    author = {Sharma, Dhruv and Purushotham, Sanjay and Reddy, Chandan K.},
    title = {{MedFuseNet: An Attention-Based Multimodal Deep Learning Model for Visual Question Answering in the Medical Domain}},
    journal = {Scientific Reports},
    volume = {11},
    number = {1},
    year = {2021},
    issn = {2045-2322}
}

@inproceedings{Xie:2020, 
author = {Xie, Yao and Chen, Melody and Kao, David and Gao, Ge and Chen, Xiang 'Anthony'},
title = {{CheXplain: Enabling Physicians to Explore and Understand Data-Driven, AI-Enabled Medical Imaging Analysis}},
year = {2020},
publisher = {Association for Computing Machinery},
booktitle = {CHI},
pages = {1–13},
numpages = {13}
}

@misc{gale2018producing,
      title={{Producing Radiologist-Quality Reports for Interpretable Artificial Intelligence}}, 
      author={William Gale and Luke Oakden-Rayner and Gustavo Carneiro and Andrew P Bradley and Lyle J Palmer},
      year={2018},
      eprint={1806.00340},
      archivePrefix={arXiv},
      primaryClass={cs.AI}
}

@Article{make3030037,
AUTHOR = {Knapič, Samanta and Malhi, Avleen and Saluja, Rohit and Främling, Kary},
TITLE = {{Explainable Artificial Intelligence for Human Decision Support System in the Medical Domain}},
JOURNAL = {Machine Learning and Knowledge Extraction},
VOLUME = {3},
YEAR = {2021},
NUMBER = {3},
PAGES = {740--770}
}

@INPROCEEDINGS {Cheng:2023,
author = {P. Cheng and L. Lin and J. Lyu and Y. Huang and W. Luo and X. Tang},
booktitle = {ICCV},
title = {{PRIOR: Prototype Representation Joint Learning from Medical Images and Reports}},
year = {2023},
volume = {},
issn = {}
}

@INPROCEEDINGS{wang2018tienet,
  author={Wang, Xiaosong and Peng, Yifan and Lu, Le and Lu, Zhiyong and Summers, Ronald M.},
  booktitle={CVF}, 
  title={{TieNet: Text-Image Embedding Network for Common Thorax Disease Classification and Reporting in Chest X-Rays}}, 
  year={2018},
  volume={},
  number={}
}

@misc{chen2024crossmodal,
      title={{Cross-Modal Causal Intervention for Medical Report Generation}}, 
      author={Weixing Chen and Yang Liu and Ce Wang and Jiarui Zhu and Shen Zhao and Guanbin Li and Cheng-Lin Liu and Liang Lin},
      year={2024},
      eprint={2303.09117},
      archivePrefix={arXiv},
      primaryClass={cs.CV}
}

@INPROCEEDINGS{Tanida:2023,
  author={Tanida, Tim and Müller, Philip and Kaissis, Georgios and Rueckert, Daniel},
  booktitle={CVPR}, 
  title={{Interactive and Explainable Region-guided Radiology Report Generation}}, 
  year={2023},
  volume={},
  number={},
  pages={7433-7442}
}

@inproceedings{Yuan:2019,
author = {Yuan, Jianbo and Liao, Haofu and Luo, Rui and Luo, Jiebo},
title = {{Automatic Radiology Report Generation Based on Multi-view Image Fusion and Medical Concept Enrichment}},
year = {2019},
booktitle = {MICCAI}
}

\end{document}